\documentclass [12pt] {article}
\usepackage{axodraw}

\catcode`@=12
\newcommand{\beq}{\begin{equation}}
\newcommand{\eeq}{\end{equation}}

\newcommand{\bea}{\begin{eqnarray}}
\newcommand{\eea}{\end{eqnarray}}

\begin{document}
\thispagestyle{empty}
\begin{flushright} UCRHEP-T354\\ ULB-TH/03-16 \\
NSF-KITP-03-32\\

May 2003\
\end{flushright}
\vspace{0.5in}
\begin{center}
{\LARGE \bf Gaugino-induced quark and lepton masses\\
in a truly minimal left-right model\\}
\vspace{1.5in}

{\bf J.-M. Fr{\`e}re$^{1,3}$ and Ernest Ma$^{2,3}$\\}
\vspace{0.2in}
{\sl $^1$ Service de Physique Th{\'e}orique, Universit{\'e} Libre de 
Bruxelles, B-1050 Brussels, Belgium\\}
{\sl $^2$ Physics Department, University of California, Riverside,
California 92521, USA\\}
{\sl $^3$ Kavli Institute for Theoretical Physics, University of California,\\}
{\sl Santa Barbara, California 93106, USA\\}
\vspace{1in}
\end{center}

\begin{abstract}
It has recently been proposed that all fundamental fermion masses (whether 
Dirac or Majorana) come from effective dimension-five operators in the 
context of a truly minimal left-right model. 
We show how a particularly economical scheme emerges in a supersymmetric 
framework, where chiral symmetry breaking originates in the gaugino sector. 
\end{abstract}
\newpage
\baselineskip 24pt

In the Standard Model of particle interactions, the spontaneous breaking of 
the $SU(2)_L \times U(1)_Y$ gauge symmetry to $U(1)_{em}$ is achieved 
through the vacuum expectation value of the scalar doublet $\Phi = (\phi^+,
\phi^0)$.  At the same time, since left-handed quarks and leptons are 
doublets under $SU(2)_L \times U(1)_Y$ whereas right-handed quarks and 
leptons are singlets, chiral symmetry is also broken by $\langle 
\phi^0 \rangle$, thus allowing quarks and leptons to acquire the usual 
Dirac masses.  The only exception is the neutrino which gets a small 
Majorana mass through the unique dimension-five operator \cite{wein,ma98}
\begin{equation}
{\cal L}_\Lambda = {f_{ij} \over 2 \Lambda} (\nu_i \phi^0 - e_i \phi^+) 
(\nu_j \phi^0 - e_j \phi^+) + H.c.
\end{equation}

Suppose we now extend the standard-model gauge symmetry to $SU(3)_C \times 
SU(2)_L \times SU(2)_R \times U(1)_{B-L}$ \cite {lr}, then the spontaneous 
breaking of $SU(2)_R \times U(1)_{B-L}$ to $U(1)_Y$ is simply achieved by 
the scalar doublet
\begin{equation}
\Phi_R = (\phi_R^+, \phi_R^0) \sim (1,1,2,1),
\end{equation}
where the notation refers to the dimension of the non-Abelian representation 
or the value of the Abelian charge $B-L$ or $Y$ in the convention
\begin{equation}
Q = I_{3L} + I_{3R} + {1 \over 2} (B-L) = I_{3L} + {Y \over 2},
\end{equation}
while the corresponding field
\begin{equation}
\Phi_L = (\phi_L^+, \phi_L^0) \sim (1,2,1,1),
\end{equation}
becomes the same as the usual scalar doublet of the Standard Model, and 
breaks $SU(2)_L \times U(1)_Y$ in turn to $U(1)_{em}$.

In other words, $\Phi_R$ and $\Phi_L$ are sufficient by themselves for the 
desired breaking of the left-right gauge symmetry all the way down to 
$U(1)_{em}$.  On the other hand, chiral symmetry of the quarks and leptons 
remains unbroken at this stage, in contrast to the case of the Standard Model. 
The conventional way to deal with this problem in the left-right context is to 
introduce a ``bidoublet'' (1,2,2,0) which, combined with either $\Phi_{R}$ 
or a similar triplet, generates both gauge-boson and fermion masses.  It is 
however also well-known that this would result in the existence of flavour 
changing neutral currents (FCNC), which must be controlled in some way.  
This solution is also particularly unpleasant if one extends the model to 
include supersymmetry.  In that case, two bidoublets (instead of one) must be 
introduced (the equivalent of 4 standard doublets) and the problem with FCNC 
is compounded.

Suppose we do not introduce this scalar bidoublet, then fermion masses are 
still possible, but they (whether Dirac or Majorana) must all come from 
dimension-five operators, as recently proposed \cite{bms}.  We would then 
have the advantage of a uniform understanding of the origin of all 
fundamental fermion masses.

The obvious (and first) proposal \cite{dwr} to generate such operators, 
outside of supersymmetry, is to introduce one massive vectorlike $SU(2)_{L} 
\times  SU(2)_{R}$ singlet fermion per observed low-energy fermion species. 
While this approach might be otherwise minimal, an easier and more economical 
mechanism exists if we extend the model to include supersymmetry.
 
Before working out the details, suppose we simply consider the supersymmetric 
extension of the scheme above, then again the quarks and leptons are 
massless without the dimension-five operators.  However, we now have the 
very interesting option of obtaining these operators radiatively through a 
special class of supersymmetry breaking terms, as was done for the 
left-handed neutrino Majorana mass in Ref.~\cite{flt}.  The resulting quark 
and lepton masses are then derived as functions of the squark and slepton 
masses together with those of the gauginos \cite{banks,ma89}.

We now turn to the minimal field content of such a model.  Under $SU(3)_C 
\times SU(2)_L \times SU(2)_R \times U(1)_{B-L}$, the quark and lepton 
superfields transform as
\begin{eqnarray}
q \sim (3,2,1,1/3), &~& q^c \sim (3^*,1,2,-1/3), \\ 
l \sim (1,2,1,-1), &~& l^c \sim (1,1,2,1),
\end{eqnarray}
where we have adopted the convention that all fermion fields are left-handed, 
with their right-handed counterparts denoted by the corresponding 
(left-handed) charge-conjugate fields.  Thus we have used here for the 
left-handed superfield $q^{c}$ the implicit notation $q^{c} = (q_{R})^{c} 
\neq (q_{L})^{c}$.  We add the following Higgs superfields:
\begin{eqnarray}
\Phi_L \sim (1,2,1,1), &~& \Phi_R \sim (1,1,2,-1), \\ 
\Phi_L^c \sim (1,2,1,-1), &~& \Phi_R^c \sim (1,1,2,1),
\end{eqnarray}
These fields allow for the obvious bilinear terms ($\Phi_L \Phi_L^c$) and 
($\Phi_R \Phi_R^c$), but also the unwanted lepton-number violating terms 
$(l_L \Phi_L)$ and $(l^c \Phi_R)$, which we must forbid by an unbroken 
discrete symmetry, i.e. all quark and lepton superfields are odd, but all 
Higgs superfields are even.  In terms of their components, $(q \Phi)$ is a 
short-hand notation for $\epsilon_{ij} q_{i} \phi_{j}$, etc.

Since we do not admit Higgs bidoublets, there are no trilinear terms in the 
superpotential of this model.  This means that all quarks and leptons remain 
massless even though $SU(3)_C \times SU(2)_L \times SU(2)_R \times U(1)_{B-L}$ 
has been broken down to $SU(3)_C \times U(1)_{em}$.  In other words, chiral 
symmetry breaking has now been separated from gauge symmetry breaking. 

Before discussing our main result, namely the generation of light fermion 
masses, we want to point out that the scalar structure of our model at low 
energy is identical to that of the Minimal Supersymmetric Standard Model 
(MSSM), once the $SU(2)_R$ fields get large masses and/or vacuum expectation 
values.  The specific FCNC problems of nonminimal left-right models are thus 
avoided.  On the other hand, FCNC could still be induced by the flavour 
structure of the supersymmetry breaking sector.
 
To generate quark and lepton masses, we must add dimension-five operators 
as proposed in Ref.~\cite{bms}, i.e. $(q \Phi_L)(q^c \Phi_R)$ for $m_u$ 
and $(q \Phi_L^c)(q^c \Phi_R^c)$ for $m_d$, etc.  As already stated, we can 
avoid here the extra fields required by the ``universal seesaw'' 
mechanism \cite{dwr}, where each quark or lepton has a heavy singlet 
counterpart.  In a supersymmetric model, no such fields are needed.  
Instead, light fermion masses may be generated by taking into account 
the presence of ``nonstandard'' (but usually present) supersymmetry breaking 
terms, as shown below.

Such terms are NOT a new addition.  In most phenomenological models, 
supersymmetry is broken in a more or less ``hidden'' sector, and characterized 
by the vacuum expectation value of some (auxiliary) field $F$. This 
hidden-sector breaking is then transferred to ``our'' world by some mediator 
(possibly gravity itself), to which we associate a mass scale $M$. This 
results in a classification of ``effective'' supersymmetry breaking terms, 
obtained by an expansion in $F/M \approx m_{susy}$ where $m_{susy}$ is 
adjusted to be the mass gap between the known particles and their 
superpartners, which we may want to keep in the 1 to 10 TeV range to 
maintain the stability of the electroweak scale.

This expansion is usually kept to lowest significant order in $1/M$, which 
yields ``superrenormalisable'' (or ``soft'') terms, like sfermion masses, 
trilinear scalar couplings or gaugino masses.  There is no reason however not 
to continue with such an expansion, other than the expected smallness of the 
effects.  Of course, since the expansion parameter is dimensional, this will 
generate ``hard'' couplings.  It should however be of little concern, since 
their impact will be small in general, unless the integration over such terms 
reaches the scale $M$, where the full theory acts as a natural cutoff.

An extensive study and classification of such terms is found in 
Ref.~\cite{martin}, with the conclusion that beyond the usual soft breakings, 
the first corrections are of order $F/M^2$ and introduce operators of the type 
$\phi \lambda \lambda$ and $\phi^{4}$ (the latter to be distinguished 
from $\phi^{*2}\phi^{2}$, which only appears at order $F^2/M^{4}$). (Here 
$\lambda$ represents a gaugino and $\phi$ the scalar part of a chiral 
superfield).  We are interested now in the quartic $\phi^{4}$ term, which 
was actually already used in a more restricted context in Ref.~\cite{flt}. 
There, the dimension-four supersymmetry breaking term $h (\epsilon_{ij} 
\tilde {l_{i}} \phi_{L,j})^2$, where $\tilde l$ and $\phi_L$ are the scalar 
partners of the corresponding superfields, is used to generate a small 
Majorana meutrino mass.  An example of such a contribution is given in Fig.~1. 
(For details, see Ref.[7]).

\begin{figure}[htb]
\begin{center}
\begin{picture}(300,160)(0,0)
\ArrowLine(50,5)(100,5)
\Line(100,5)(200,5)
\ArrowLine(250,5)(200,5)
\DashArrowLine(100,5)(150,90)5
\DashArrowLine(200,5)(150,90)5
\DashArrowLine(125,132)(150,90)5
\DashArrowLine(175,132)(150,90)5
\Text(75,-5)[]{$\nu$}
\Text(225,-5)[]{$\nu$}
\Text(116,140)[]{$\langle \phi_L^0 \rangle$}
\Text(186,140)[]{$\langle \phi_L^0 \rangle$}
\Text(116,48)[]{$\tilde \nu$}
\Text(186,48)[]{$\tilde \nu$}
\Text(150,-8)[]{$\tilde z$}
\Text(150,5)[]{$\times$}
\end{picture}
\end{center}
\caption{Example of a one-loop generation of Majorana neutrino mass.}
\end{figure}
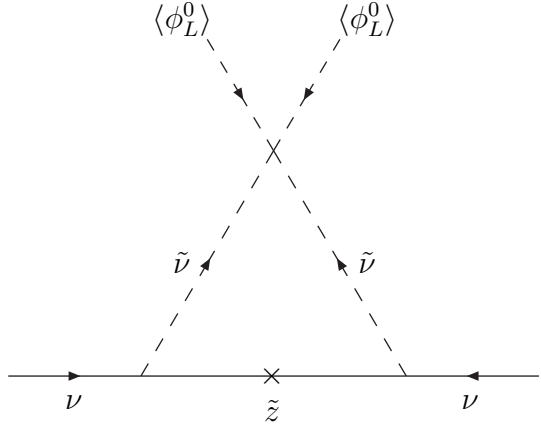

This graph shows clearly how we differ from the canonical seesaw mechanism. 
The quantum numbers of the external fermions are carried through the graph 
by the sfermions (eliminating the need for heavy Majorana particles or 
heavy vectorlike partners), while the chiral symmetry breaking is provided 
in a (largely) universal way by the gauginos.  Of course, the effect 
initially evaluated for left-handed neutrinos is tiny (as desired), but 
in the present left-right extension with operators such as $ h (\tilde q 
\Phi_L)(\tilde q^c \Phi_R)$, etc., a Dirac mass would be enhanced by the 
large ratio of $\langle \phi^0_{R} \rangle=v_{R} / \sqrt{2}$ to $\langle 
\phi^0_{L} \rangle=v_{L}/ \sqrt{2}$., implying that $SU(2)_R$ should be 
broken at a very high scale.

For the charged leptons, the graphs to be considered are similar to Fig.~1, 
but with both $SU(2)_L$ and $SU(2)_R$ fields involved, and the chirality 
flip is due again to the exchange of a neutralino (typically the partner of 
the light $U(1)_Y$ gauge field, usually named the bino ($\tilde B$), the 
only neutral gaugino to have both significant $SU(2)_L$ and $SU(2)_R$ 
couplings). In the case of quarks, both neutralino and gluino exchange must 
be considered, the latter providing the dominant contribution 
\cite{banks,ma89}, as shown in Fig.~2.

\begin{figure}[htb]
\begin{center}
\begin{picture}(300,160)(0,0)
\ArrowLine(50,5)(100,5)
\Line(100,5)(200,5)
\ArrowLine(250,5)(200,5)
\DashArrowLine(100,5)(150,90)5
\DashArrowLine(200,5)(150,90)5
\DashArrowLine(125,132)(150,90)5
\DashArrowLine(175,132)(150,90)5
\Text(75,-5)[]{$q$}
\Text(225,-5)[]{$q^c$}
\Text(116,140)[]{$\langle \phi_L^0 \rangle$}
\Text(186,140)[]{$\langle \phi_R^0 \rangle$}
\Text(116,48)[]{$\tilde q$}
\Text(186,48)[]{$\tilde q^c$}
\Text(150,-8)[]{$\tilde g$}
\Text(150,5)[]{$\times$}
\end{picture}
\end{center}
\caption{One-loop generation of quark mass.}
\end{figure}
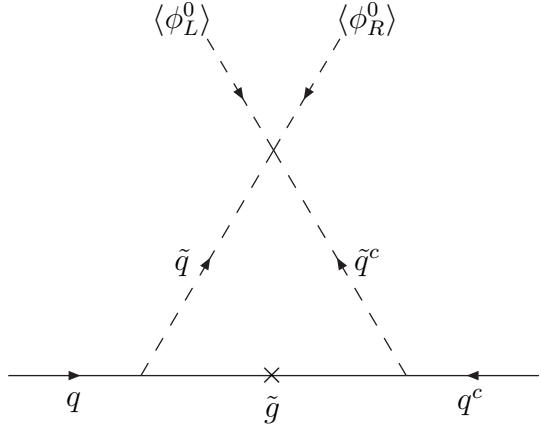

The new term $v_{L} v_{R}$ induces mixing between $SU(2)_L$ and $SU(2)_R$ 
sfermions, and we should in principle diagonalize this matrix and deal with 
the exchanges of both eigenstates (see for instance Ref.~\cite{ma89}). 
However, excluding the case of $m_t$ (to be discussed later), the mixing 
term is always safely small, and the 
evaluation of the triangle graph is accurate enough.  In the following 
we assume that all superpartner masses (gauginos, squarks, and sleptons) 
are similar and denoted by $\tilde{m}$, whereas $v_L$ refers to either of 
the 2 scalar doublet vacuum expectation values of the MSSM.  We obtain in 
this way for the gluino contribution to the quark mass the order of 
magnitude \begin{equation}
m_{q} \simeq {1 \over 32 \pi^{2}} \left( {4 \over 3} \right) \left( 
{2g_{s}^{2} \over \tilde m} \right) \left( { h v_{L} v_{R} \over 2} \right),
\end{equation}
where we have used the quartic coupling $h (\tilde q \Phi_{L})(\tilde q^{c} 
\Phi_{R})$.  Recalling our previous discussion on the origin of $h$, we set 
it equal to $\kappa \,\tilde{m} / M$ with $M$ the ``mediator'' mass and obtain
\begin{equation}
m_{q} \simeq {\alpha_{s} \kappa \,v_{L} \over 6 \pi} \left( {v_{R} \over M} 
\right).
\end{equation}
Note that the scale $\tilde{m}$ disappears from this equation.  However 
there are indirect constraints which we must consider.  Nevertheless, 
let us ignore these constraints for the moment and proceed to determine the 
necessary conditions for the radiative generation of realistic quark masses, 
the strongest requirement obviously coming from the top quark. 
Since $\kappa$ is expected to be of order unity or less, it is readily seen 
that realistic quark (and lepton) masses require a large $v_{R}$.  For the 
top quark mass, $v_R/M \sim 10^2$ would be required.  In ``gravity-mediated'' 
schemes, $M$ is already of order the Planck mass, which tells us that 
$v_R > M$ is not a reasonable choice.  On the other hand, if $M$ is smaller, 
as in gauge-mediated supersymmetry breaking, $v_R > M$ is possible, but then 
$v_R > \sqrt F$  as well.  In other words, $SU(2)_R$ breaking must occur 
without supersymmetry breaking.  For example, if we choose $v_R \sim 10^{12}$ 
GeV for leptogenesis, then $M \sim 10^{10}$ GeV and $\sqrt F \sim 10^7$ GeV 
for $m_{susy} \sim 10$ TeV.  At the same time, using the analog of Eq.~(10), 
we obtain $m_\nu \sim 0.1$ eV, which is just right for neutrino 
oscillations.

The large $(v_R/M)$ ratio also raises a strong concern that the expansion in 
powers of $1/M$ may not be valid. This would certainly be true unless the 
connection between $v_R$ and the low-energy sector is 
systematically suppressed, such that the true expansion parameter is not 
$v_R/M$ but a smaller combination, as for example in the graph 
studied here.  Whether or not this is actually realized 
depends on the (unknown) specific supersymmetry breaking mechanism.
Imposing this suppression of the coupling has immediate consequences. 
Maintaining the supersymmetry 
mass scale $\tilde{m}$ in a reasonable range (say 1 to 10 TeV) requires that 
a large value of $v_R$ does not generate excessive off-diagonal corrections 
to the diagonal squark masses.  This contribution is of course also given by 
the vertex already considered, and is of order $\kappa (v_L / \tilde{m})
(v_R / M)(\tilde{m})^2$.  To keep this contribution of order $(\tilde{m})^2$, 
we require $\tilde{m} \sim v_L ( v_R / M)$ and thus a rather large 
supersymmetry mass scale (of order 10 TeV).   A potentially even stronger 
constraint comes from the quartic coupling $(\tilde q^{c} \Phi_{R})(\tilde 
q^{c} \Phi_{R})^\dag$.  Such a term is generically of order $F^2 / M^4$ 
(see Ref.~\cite{martin}) and generates a correction of order $(F/M)^2 
(v_R/M)^2 \sim \tilde{m}^2 (v_R /M)^2 $.  Keeping this correction under 
control with a large value of $v_R/M$ implies that the coefficient of this 
term is for some reason strongly suppressed or perhaps even absent.  

If instead $v_R \sim M$ is assumed, all quark and lepton masses may still be 
obtained from Eq.~(10) except for $m_t$.  In that case, we may invoke the 
dynamical mechanism of Bardeen, Hill, and Lindner (BHL) \cite{bhl} which also 
uses a higher-dimensional operator.  [In the case of supersymmetry, this 
mechanism has also been discussed \cite{ccwbs}.]  Since there are already 
fundamental scalars in supersymmetry which can share the vacuum expectation 
values of electroweak 
symmetry breaking, the correct $m_t$ from the BHL mechanism may also be 
obtained. [In the Standard Model, the simplest version of the BHL model 
predicts $m_t = 226$ GeV for a cutoff of $10^{16}$ GeV and even a bigger 
value for a lower cutoff.]  As far as the low-energy phenomenology is 
concerned (at energies up to the supersymmetry scale $\tilde{m}$), the 
effective Yukawa couplings between light fermions and physical scalar 
bosons are then modified with respect to the MSSM, because there would be 
an additional effective BHL Higgs doublet.

We should also point out that in the present scheme, the chiral symmetry 
breaking can be traced first to the gaugino sector, and through it and the 
mediators, to the supersymmetry breaking scenario itself.  In such 
circumstances, one might expect some form of ``superlight axion mechanism'' 
to take place, associating the axion primarily to the first source of chiral 
symmetry breaking, i.e. the dynamical phase of the gluino as proposed 
in Ref.~\cite{axion}.  The details depend of course on the choice of a 
definite model for the supersymmetry breaking mechanism.

In conclusion, we have proposed a supersymmetric left-right model, where 
the gauge symmetry breaking is achieved simply by 4 doublet superfields, 
without breaking the chiral symmetry of the quarks and leptons.  The latter 
is accomplished instead by supersymmetry breaking terms involving 
gauginos, sfermions, and a quartic scalar interaction.  We have thus a 
uniform understanding of all fermion masses (except possibly for $m_t$): 
they are radiatively generated by supersymmetry breaking and Dirac masses 
are much larger than Majorana masses because $v_R >> v_L$.\\[5pt]

This work was supported in part by the U.~S.~Department of Energy
under Grant No.~DE-FG03-94ER40837, by the French Community of Belgium 
under contracts IISN and ARC, by the Belgian federal government under 
IAP 5/27, and by the U.~S.~National Science Foundation under Grant 
No.~PHY99-07949.  In this last respect, the authors thank the KITP 
(U.~C.~Santa Barbara) for its gracious hospitality and friendly atmosphere, 
and the organisers of its 2003 Neutrino Workshop, which inspired this work.

\newpage
\bibliographystyle{unsrt}

\begin{thebibliography}{99}
\bibitem{wein} S. Weinberg, Phys. Rev. Lett. {\bf 43}, 1566 (1979).
\bibitem{ma98} E. Ma, Phys. Rev. Lett. {\bf 81}, 1171 (1998).
\bibitem{lr} J. C. Pati and A. Salam, Phys. Rev. {\bf D10}, 275 (1974).
\bibitem{bms} B. Brahmachari, E. Ma, and U. Sarkar, Phys. Rev. Lett. {\bf 91}, 
011801 (2003).
\bibitem{dwr} A. Davidson and K. C. Wali, Phys. Rev. Lett. {\bf 59}, 393,
(1987); S. Rajpoot, Phys. Rev. {\bf D36}, 1479 (1987).
\bibitem{martin} S. P. Martin, Phys. Rev. {\bf D61} 035004 (2000).
\bibitem{flt} J.-M. Fr{\`e}re, M. V. Libanov, and S. V. Troitsky, Phys. Lett. {\bf B479}, 343 (2000).
\bibitem{banks} T. Banks, Nucl. Phys. {\bf B303}, 172 (1988).
\bibitem{ma89} E. Ma, Phys. Rev. {\bf D39}, 1922 (1989).
\bibitem{bhl} W. A. Bardeen, C. T. Hill, and M. Lindner, Phys. Rev. {\bf D41}, 
1647 (1990).
\bibitem{ccwbs} M. Carena, T. E. Clark, C. E. M. Wagner, W. A. Bardeen, and 
K. Sasaki, Nucl. Phys. {\bf B369}, 33 (1992).
\bibitem{axion} D. A. Demir and E. Ma, Phys. Rev. {\bf D62}, 111901(R) (2000).

\end{thebibliography}

\end{document}